\documentclass[twocolumn,showpacs,preprintnumbers,amsmath,amssymb,APS]{revtex4}
\usepackage{graphicx}
\usepackage{bm}

\newcommand{\ud}{\mathrm{d}}
\newcommand{\cref}[1]{(\ref{#1})}

\begin{document}

\title{Spontaneous Charging and Crystallization of Water Droplets in Oil}

\author{Joost de Graaf}%
 \email{j.degraaf@phys.uu.nl}
\author{Jos Zwanikken}%
\author{Markus Bier}%
\author{Arjen Baarsma}%
\author{Yasha Oloumi}%
\author{Mischa Spelt}%
\author{Ren\'e van Roij}%
 \email{r.vanroij@uu.nl}
\affiliation{%
Institute for Theoretical Physics, Utrecht
University, \\ Leuvenlaan 4, 3584 CE Utrecht, The Netherlands
}%

\date{\today}

\begin{abstract}
We study the spontaneous charging and the crystallization of
spherical micron-sized water-droplets dispersed in oil by
numerically solving, within a Poisson-Boltzmann theory in the
geometry of a spherical cell, for the density profiles of the
cations and anions in the system. We take into account screening,
ionic Born self-energy differences between oil and water, and
partitioning of ions over the two media. We find that the surface
charge density of the droplet as induced by the ion partitioning
is significantly affected by the droplet curvature and by the
finite density of the droplets. We also find that the salt
concentration and the dielectric constant regime in which
crystallization of the water droplets is predicted is enhanced
substantially compared to results based on the planar oil-water
interface, thereby improving quantitative agreement with recent
experiments.
\end{abstract}


\maketitle

\section{\label{sec:intro}Introduction}

It is well-established that water and oil do not mix: droplets of
water in oil (or droplets of oil in water) tend to coalesce such
that the oil-water mixture coarsens until macroscopic phase
separation of oil and water is achieved. It is also
well-established that this coarsening process can be delayed or
even prevented by additives such as surfactants or colloidal
particles, which adsorb to the oil-water interface and thereby
stabilize the droplets, either thermodynamically (such as in
micro-emulsions) or kinetically (such as in Pickering
emulsions)~\cite{pickart,binks}. Recently, however, experimental
observations by Leunissen and coworkers~\cite{leuniss,miriaml}
revealed stable micron-sized water droplets in somewhat polar oils
\emph{without} any additives. In fact, under appropriate
conditions the oil-dispersed water droplets could even form a
body-centered cubic (bcc) crystalline phase with a lattice spacing
of the order of 10 $\mu$m, and some of these crystals have been
stable for almost two years now without any observation of droplet
coalescence~\cite{miriam}. The mechanism by which these water
droplets are stabilized was argued to stem from an asymmetric
\emph{partitioning} of the (ever-present) monovalent cations and
anions over the oil and water phase; in the experiments of
Refs.~\cite{leuniss,miriaml} with the oil cyclohexyl bromide the
ions involved include $\mathrm{H}^+$ and $\mathrm{Br}^-$, and the
somewhat larger water affinity of the former compared to the
latter should lead to positively charged water droplets. This
suggested mechanism was confirmed in theoretical
calculations~\cite{miriaml,zwanikk} of monovalent ions in the
vicinity of a \emph{planar} oil-water interface, on the basis of
Poisson-Boltzmann theory combined with ionic Born self energy in
water and oil~\cite{born,verwey,israel}. Although more advanced
models could be invoked, e.g., involving more ionic
correlations~\cite{pratt,Shk0,Shk1,gros} or a better account of
the ionic self-energy \cite{onuki0,onuki1,onuki2,netz}, the
relatively simple model used in Refs.~\cite{miriaml,zwanikk}
showed surface charge densities of the order of $10-100$
elementary charges per $\mu\mathrm{m}^{2}$ and hence $100-1000$
charges for micron-sized water-droplets. With such a droplet
charge the observed bcc crystals of water-droplets in oil could be
explained, at least qualitatively, and therefore we use this
relatively simple Poisson-Boltzmann-Born model for further
theoretical explorations.

The theory presented in Refs.~\cite{miriaml,zwanikk} considers a
\emph{planar} oil-water interface separating two half spaces of
oil and water. The advantage of this assumption lies in the fact
that it allows for some analytic expressions for the surface
charge density and the ionic contribution to the interfacial
tension within nonlinear Poisson-Boltzmann theory~\cite{kung},
which leads to an efficient scheme to analyze the parameter space.
However, one could \emph{a priori} expect quantitative
shortcomings due to the assumed planar geometry, e.g., because the
typical experimental droplet radius of about 1 $\mu$m is quite a
bit smaller than the typical screening length of about 10 $\mu$m
in the oil phase, or because the typical lattice spacing in the
crystal is of the same order as the screening length such that the
net charge of the droplets could be affected by the nearby other
droplets. In order to investigate these effects we extend in this
paper the theory of Refs.~\cite{miriaml,zwanikk} from the planar
to the spherical geometry. This will be done in the context of a
cell model~\cite{alex}, where a single spherical droplet is
considered in the center of a spherical cell with a finite volume
representing the density of droplets.  Whereas this geometry does
no longer allow for analytic solutions of the Poisson-Boltzmann
equation, the numerical solution is, however, fairly
straightforward because of the radial symmetry. This geometry
therefore enables us to study simultaneously the effects of
droplet-curvature and droplet-density on the ionic double layers
and on preferential adsorption and charging in the vicinity of the
droplet surface. We will show that these effects, when compared to
the planar limit results, give rise to a significantly larger
crystallization regime for water-droplets in oil (due to a larger
surface charge), and to a much smaller surface charge for
oil-droplets in water. In fact, our numerical predictions for the
crystallization regime are now quantitatively closer to the
experimentally observed one, although there is still some
deviation that we attribute to other shortcomings and
oversimplifications of our microscopic model, e.g., the crude
approximation of describing the ionic self-energies in oil and
water by a simple Born energy. Our present results indicate,
however, that the essential physical mechanism of preferential ion
partitioning can indeed explain the crystallization of
water-droplets in oil as observed in Refs.~\cite{leuniss,miriaml}.

\section{\label{sec:DFT}Density Functional Theory\protect\\ for Saline Emulsions}

\subsection{\label{sub:WS}Wigner-Seitz cell approach}

\begin{figure}
\includegraphics[width=2.5in]{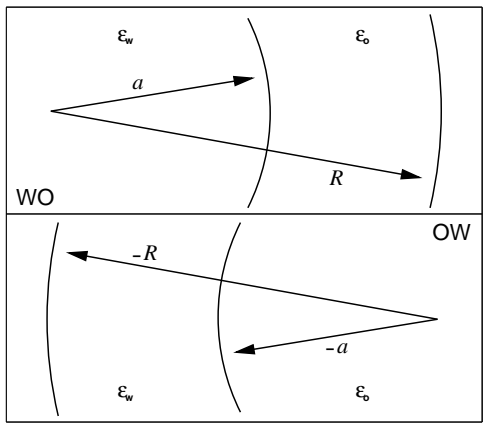}
\caption{\label{fig:config} A single Wigner-Seitz cell with radius
$R$ centered on a droplet of radius $a$. To distinguish between a
water-in-oil emulsion (WO,$+$) and an oil-in-water (OW,$-$)
emulsion, a sign convention has been introduced.}
\end{figure}

We consider an emulsion of water-in-oil droplets (WO) of total
volume $V$ containing $N$ identical droplets with radius $a$. The
volume of water in the system is defined as $x V \equiv 4 \pi N
a^{3}/3$, with $x$ the volume fraction, hence $(1-x)V$ is the
volume of oil. The theory describing emulsions of oil-droplets in
water (OW) is analogous to that for WO systems outlined in this
section. The emulsion contains monovalent ions with ionic radii
$a_{\pm}$, which are typically $2-4$ {\AA}. Using a cell
model~\cite{alex}, we reduce the $N$-droplet problem to that of a
single droplet in a spherically symmetric Wigner-Seitz cell, see
Fig.~\ref{fig:config}, with radius $R \equiv a x^{-1/3}$, such
that $V = 4 \pi N R^{3}/3$.

In this single cell we consider a spherical oil-water interface
located at $r=a$. The interface separates two bulk phases
consisting of water $(0 < r < a)$ and oil $(a < r < R)$. Both, oil
and water, are considered to be incompressible linear dielectrics,
which means that the solvent background is characterized by the
relative dielectric constant $\epsilon_{w}$ and $\epsilon_{o}$, respectively, see
Fig.~\ref{fig:config}. The dielectric profile (relative to the
dielectric constant of vacuum $\epsilon_{v}$) is a step-function
$\epsilon(r) = \epsilon_{w}$ if $0 < r < a$ and $\epsilon(r) =
\epsilon_{o}$ if $a < r < R$. The ions are described by
spherically symmetric ionic density profiles $\rho_{\pm}(r)$. The
ions experience Coulombic ion-ion interactions, which we treat in
a mean-field fashion, and ion-medium interactions.

The ion-medium interaction is taken into account via an external
potential acting on the ions. Due to the dielectric properties of
oil and water, the ions have different electrostatic self-energies
in the two solvents. Using the Born approximation \cite{israel},
this self-energy of a cation ($+$) and an anion ($-$) is given by
$E_{\pm}(\epsilon_{i}) \equiv
e^{2}/(2\epsilon_{v}\epsilon_{i}a_{\pm})$, with $e$ the elementary
charge and $i = w,o$. This self-energy and the above dielectric
profile $\epsilon(r)$, allows us to rewrite the external potential
acting on a cation and an anion as $V_{\pm}(r) =
E_{\pm}(\epsilon(r)) - E_{\pm}(\epsilon_{w})$, where we note that
this potential is constructed to be zero in water. For realistic
$\epsilon_{o} \approx 4-20$ the potential is of order
$(1-20)k_{B}T$ in oil, i.e., the ions prefer to be in the water.
We also use the notation $V_{\pm}(r) = 0$ if $0 < r < a$ and
$V_{\pm}(r) = k_{B} T f_{\pm}$ if $a < r < R$, with $f_{\pm}$
dimensionless and implicitly dependent on $\epsilon_{w}$ and
$\epsilon_{o}$. Here $k_{B}T$ is the thermal energy, $k_{B}$
Boltzmann's constant, and $T$ the temperature.

\subsection{\label{sub:PBE}Poisson-Boltzmann equation}

Using the above external potential we employ the framework of
density functional theory~\cite{evans0,evans1} to calculate the
equilibrium density profiles $\rho_{\pm}(r)$. The grand-potential
functional $\Omega[\rho_{\pm}]$ for a single Wigner-Seitz cell can
be written as
\begin{eqnarray}
\nonumber \beta \Omega[\rho_{\pm}] & = & 4 \pi \sum_{i = \pm}
\int_{0}^{R} r^{2} \rho_{i}(r) \left[
\log\left(\frac{\rho_{i}(r)}{\rho_{s}}\right) - 1  \right. \\
\label{eq:GPnoneq} & & \left. + \frac{1}{2} q_{i}
\phi(r,[\rho_{\pm}]) + \beta V_{i}(r)
 \right] \ud r ,
\end{eqnarray}
with $\beta = 1/(k_{B}T)$ and the ionic valencies $q_{\pm} = \pm
1$. The first line is the ideal-gas grand-potential functional.
The second line describes the ion-ion Coulomb interaction in
mean-field approximation and the ion-solvent interactions
characterized by the external fields. The chemical potentials are
represented in the form of an ion concentration $\rho_{s}$, which
is actually the ion concentration in a water reservoir in
equilibrium with the emulsion. The electrostatic interactions
between the ions in Eq.~\cref{eq:GPnoneq} are given in terms of
the electrostatic potential functional $k_{B} T
\phi(r,[\rho_{\pm}])/e$, which satisfies the Poisson equation
\begin{equation}
\label{eq:poisprim} \epsilon_{v} \epsilon(r) \nabla^{2}
\phi(r,[\rho_{\pm}]) = -4 \pi \beta e^{2} \sum_{i = \pm}q_{i}
\rho_{i}(r),
\end{equation}
with boundary conditions
\begin{eqnarray}
\label{eq:BC1} \lim_{r \uparrow a} \epsilon_{w} \phi'(r,[\rho_{\pm}]) & = & \lim_{r\downarrow a} \epsilon_{o} \phi'(r,[\rho_{\pm}]); \\
\label{eq:BC2} \lim_{r\downarrow 0}\phi'(r,[\rho_{\pm}]) & = &
\lim_{r\uparrow R}\phi'(r,[\rho_{\pm}]) = 0,
\end{eqnarray}
where the prime denotes a derivative w.r.t. $r$.

Minimizing the grand-potential functional leads to the
Euler-Lagrange equation $\delta \Omega / \delta \rho_{\pm}(r) =
0$, which can be rewritten as Boltzmann distributions
\begin{equation}
\label{eq:rho} \rho_{\pm}(r) = \rho_{s}\exp( -\beta V_{\pm}(r) \mp
\phi(r,[\rho_{\pm}]) ).
\end{equation}
In practice we implement the condition that $\rho_{\pm}(0) =
\rho_{s}$ for WO emulsions and $\rho_{\pm}(-R) = \rho_{s}$ for OW
emulsions, where we adhere to the sign convention explained in
Fig.~\ref{fig:config}. In the systems studied $\kappa_{w} a \gg
1$, with $\kappa^{-1}_{w}$ the Debye length in water, so that the
water phase can indeed be considered a bulk phase and hence acts as a
salt reservoir with total ion concentration $2\rho_{s}$. Using
Eq.~\cref{eq:rho} the Poisson equation reduces to
\begin{equation}
\label{eq:Poisson} \nabla^{2} \phi(r) =
\kappa(r)^{2}\sinh(\phi(r)-\phi_{c}(r)) \quad \ (r \ne a),
\end{equation}
where we have introduced $\kappa(r) = \kappa_{w}$ if $0 < r < a$
and $\kappa(r) = \kappa_{o}$ if $a < r < R$, with $\kappa_{i}^{2}
\equiv 8\pi \beta e^{2} \rho_{i}/(\epsilon_{v}\epsilon_{i})$ in
medium $i = o,w$ with $\rho_{w} = \rho_{s}$ and $\rho_{o} =
\rho_{s}\exp(-[f_{+} + f_{-}]/2)$ the bulk ion concentrations. The
Donnan potential $\phi_{c}(r) = \beta[V_{-}(r) - V_{+}(r)]/2$
follows from the local charge neutrality in the bulk liquids. Note
that we have dropped the explicit $[\rho_{\pm}]$ dependence from
$\phi(r,[\rho_{\pm}])$ in Eq.~\cref{eq:Poisson}, since the above differential equation is
not explicitly $\rho_{\pm}$ dependent. Equation~\cref{eq:Poisson}
together with its boundary conditions (see Eqs.~\cref{eq:BC1}
and~\cref{eq:BC2}) has a unique solution and can be solved
numerically on an $r$-grid by employing standard numerical
algorithms. Typically we require several thousand non-equidistant
grid points, with a relatively small grid spacing close to $r =
a$, tailored to the screening length in the oil and water phase,
respectively.

\subsection{\label{sub:CHST}Ion-induced physical quantities}

Using the numerical solution for $\phi(r)$ (and hence for
$\rho_{\pm}(r)$) we can determine the charge and the excess
interfacial tension of the droplet, and the inter-droplet coupling
parameter. The number of net unit charges induced by ion
partitioning on the droplet is given by
\begin{equation}
\label{eq:charge} Z = 4 \pi \sum_{i = \pm} \int_{0}^{a} r^{2}
q_{i} \rho_{i}(r) \ud r.
\end{equation}

Two charged water droplets, at separation $r> 2a$, are assumed to
interact with each other through a screened Coulomb
droplet-droplet interaction potential
\begin{equation}
\label{eq:coulomb} U(r) = \frac{Z^{2} e^{2} }{\epsilon_{v}
\epsilon_{o} }\frac{\exp(\kappa_{o}(2a -
r))}{(1+\kappa_{o}a)^{2}r}.
\end{equation}
We introduce the \emph{coupling parameter}
\begin{equation}
\label{eq:GAMMA} \Gamma \equiv \beta U(\rho^{-1/3}) ( 1 + k +
k^{2}/2),
\end{equation}
with $k = \kappa_{o} \rho^{-1/3}$ and $\rho^{-1/3} = (4
\pi/3)^{1/3}R$, i.e., $\rho = N/V$ the droplet density. It is
empirically known from point-Yukawa ($\kappa_{o}a = 0$)
simulations \cite{plsmpr1} that crystallization occurs when
$\Gamma >106$. Even though the systems of present interest do have
a finite `hard' core, we can still apply this freezing criterion,
because $U(2a) \gg k_{B}T$ and $\kappa_{o}a \lesssim 1$ for most
of our parameters, i.e., the repulsions are dominated by the
screened-Coulomb part rather than the hard core.

The \emph{excess interfacial tension} $\gamma$ induced by ionic
partitioning is defined as $\gamma \equiv (\Omega[\rho_{\pm}] -
\Omega[\rho_{h}])/(4\pi a^{2})$, where $\rho_{h}(r) = \rho_{w}$ if
$0 < r < a$ and $\rho_{h}(r) = \rho_{o}$ if $a < r < R$, i.e., the
difference between the grand-potential of the system and the
grand-potential of a homogeneous reference system (per area).
Using the above definition, $\gamma$ can be rewritten as
\begin{equation}
\label{eq:gamma} \beta \gamma  =  -\frac{1}{a^{2}} \sum_{i = \pm}
\int_{0}^{R} r^{2} \left[\rho_{i}(r) - \rho_{h}(r)
\vphantom{\frac{1}{2}} + \frac{1}{2} q_{i} \rho_{i}(r) \phi(r)
\right] \ud r .
\end{equation}
The term `excess' refers to the fact that the total interfacial
tension reads $\gamma_{\mathrm{tot}} = \gamma_{\mathrm{bare}} +
\gamma$, where $\gamma_{\mathrm{bare}}$ is the bare oil-water
interfacial tension, i.e., for an oil-water interface without the
presence of ions. This bare interfacial tension is positive and
typically of the order $1-10$ mN/m, whereas $\gamma$ turns out to
be negative and of the order $10 - 100$ nN/m.

\section{\label{sec:results}Results and Discussion}

\subsection{\label{sub:parsel}Choice of parameters}

From the experiments of Refs.~\cite{leuniss,miriaml} we know that
crystallization of water-in-oil droplets has been observed for the
following parameters. Water has $\epsilon_{w} = 80.0$ at $T = 293$
K. For water-droplets in CHB (cyclohexyl bromide) we find from an
analysis of the published snapshots that $a = 1.1 \pm 0.2$ $\mu$m
and $l = 9.6 \pm 0.8$ $\mu$m, with $l$ the nearest neighbor
distance in a bcc crystal~\cite{leuniss}. Elementary geometry
shows that the volume fraction of water is then given by $x = \pi
\sqrt{3} (a/l)^{3}$, which yields $x = (8.1 \pm 1.6)\cdot10^{-3}$.
For CHB $\epsilon_{o} = 7.9$ and electro-conductivity measurements
give an indication of the bulk salt concentrations $\rho_{w}$ and
$\rho_{o}$~\cite{leuniss}. According to Ref.~\cite{miriaml} the
major constituent ions are H$^{+}$, OH$^{-}$ and
$\mathrm{Br}^{-}$, with $a_{\mathrm{H}^{+}} = 2.8$ {\AA} and
$a_{\mathrm{Br}^{-}} = 3.3$ {\AA}~\cite{leuniss}. Likewise, for
water-droplets in a CHB-decalin mixture (see Ref.~\cite{leuniss}
for details) we find $a = 1.3 \pm 0.2$ $\mu$m, $l = 16 \pm 2$
$\mu$m and hence $x = (2.9 \pm 0.6)\cdot10^{-3}$. This CHB-decalin
mixture has $\epsilon_{o} = 5.6$ and $\kappa_{o}^{-1} \gtrsim 3.6$
$\mu$m~\cite{miriaml}. Again the contributing ions are
$\mathrm{H}^{+}$, $\mathrm{OH}^{-}$ and $\mathrm{Br}^{-}$, but
their respective concentrations in CHB-decalin or water have a
high degree of uncertainty.

In our theoretical investigation we have chosen system parameters
in the range indicated by the experiments of Ref.~\cite{leuniss}.
However, to capture the physics of the curvature effects present
at spherical interfaces we do not fully take into account the
complex chemistry described above. Our \emph{basis parameter set}
is $R = 10$ $\mu$m, $a = 1$ $\mu$m, $\rho_{w} = 10^{-3}$ M,
$\epsilon_{w} = 80$, $\epsilon_{o} = 5$, $a_{+} = 3.6$ {\AA} and
$a_{-} = 3.0$ {\AA}. We vary one or more of these parameters at a
time and examine the effect on the physical quantities $Z$,
$\gamma$, and $\Gamma$. For this basis parameter set we have
$f_{+} = 14.8$, $f_{-} = 17.8$ (using the Born approximation), $x
= 10^{-3}$, $\rho_{o} = 8.15\cdot10^{-11}$ M, $\kappa_{w}^{-1} =
9.63$ nm and $\kappa_{o}^{-1} = 8.44$ $\mu$m. We take two ionic
species for simplicity and numerical convenience. The positive
ionic radius corresponds to that of $\mathrm{Na}^{+}$ and the
negative to that of $\mathrm{Cl}^{-}$~\cite{ionic1,ionic2,ionic3}.
At this point the choice for the ion concentration in water seems
a bit arbitrary, but we will show that it is in fact reasonable.

The results for spherical interfaces have been calculated using
numerical techniques, whereas in the planar limit results can be
determined analytically \cite{kung}. Note that the planar limit
describes flat oil-water interfaces, therefore the planar system
should correspond to a spherical Wigner-Seitz cell with $a, R-a
\gg \kappa_{o}^{-1}$, i.e., the interface is locally flat on the
scale of the Debye length of oil; we will denote this limit as
$a,R \rightarrow \infty$.

\subsection{\label{sub:init}Preliminary analysis}

\begin{figure}
\includegraphics[width=3.375in]{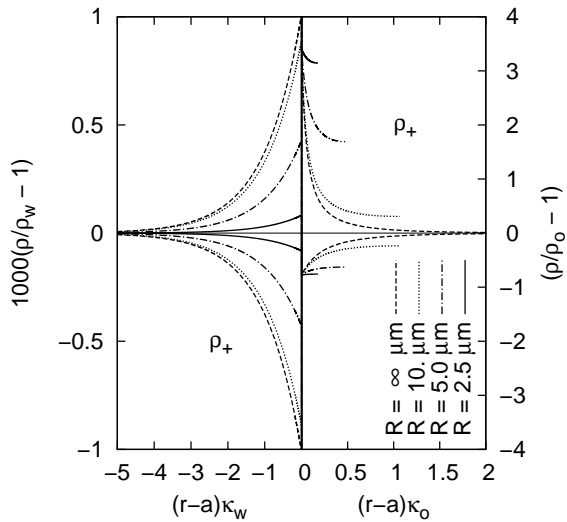}
\caption{\label{fig:prefer} The double-layer near the interface of
a spherical water-in-oil droplet, showing ion partitioning of the
ions. Here $a = 1$ $\mu$m, $\rho_{w} = 10^{-3}$ M, $\epsilon_{w} =
80$, $\epsilon_{o} = 5$, $a_{+} = 3.6$ {\AA} and $a_{-} = 3.0$
{\AA}, which gives $\kappa_{w}^{-1} = 9.63$ nm, $\kappa_{o}^{-1} =
8.44$ $\mu$m and $\rho_{o} = 8.15\cdot10^{-11}$ M. The deviation
from the homogeneous density profile is given as a function of the
distance from the interface measured in screening lengths, for
several values of the Wigner-Seitz cell radius $R$. The upper-left
and lower-right quadrants correspond to $\rho_{-}(r)$ profiles,
whereas the lower-left and upper-right correspond to the
$\rho_{+}(r)$ profiles. Note that the lines on the oil side of the
interface terminate at $(R-a)\kappa_{o}$, i.e., $r=R$.}
\end{figure}
Figure.~\ref{fig:prefer} shows the ionic density profiles for an
$a = 1$ $\mu$m water-in-oil droplet near the interface for several
Wigner-Seitz cell radii $R$. Note the different scaling of all
\emph{four} axes, and that $x$ increases from $0$ ($R \rightarrow
\infty$), to $0.001$ ($R = 10$ $\mu$m), to $0.008$ ($R = 5$
$\mu$m), to $0.064$ ($R = 2.5$ $\mu$m). Ion partitioning causes
the water phase to become negatively charged for $a_{+} > a_{-}$,
whereas the oil phase picks up an equal but opposite charge. Note
that local charge neutrality at $r = R$ is violated for the finite
Wigner-Seitz cells. This finite cell `compresses' the double-layer
inside the droplet w.r.t. that of the planar limit system,
resulting in a droplet charge reduction, as can be seen from the
shrinkage of the area enclosed by $\rho_{+}(r)$ and $\rho_{-}(r)$
at the water side in Fig.~\ref{fig:prefer}. Of course, the same
area-shrinkage occurs in the oil phase by the condition of global
charge neutrality, however, this effect is not clearly visible.
\begin{table}
\caption{\label{tab:diff}Physical quantities corresponding to
Fig.~\ref{fig:prefer} compared to those calculated analytically
for a planar system.}
\begin{ruledtabular}
\begin{tabular}{cccccc}
$a$ & $R$ & $\sigma$\footnote{$\sigma \equiv Z/(4 \pi a^{2})$} & $Z$ & $\gamma$ & $\Gamma$\\
($\mu$m) & ($\mu$m) & (e/$\mu$m$^{2}$) & (e) & nN/m & -\\
\hline
$\infty$ & $\infty$\footnote{A planar system, where only $\sigma$ and $\gamma$ can be determined.} & -1.34 & -$\infty$ & -3.7 & - \\
1.0 & $\infty$ & -11.61 & -145.9 & -34.8 & 0.0 \\
1.0 & 10.0 & -10.23 & -128.5 & -30.6 & 16.8 \\
1.0 & 5.0 & -4.93 & -61.9 & -14.0 & 8.7 \\
1.0 & 2.5 & -0.94 & -11.8 & -2.4 & 0.64 \\
\end{tabular}
\end{ruledtabular}
\end{table}
The value of the physical quantities for the various systems in
Fig.~\ref{fig:prefer} is shown in Table~\ref{tab:diff} together
with the values for the corresponding planar system. Note that we
have introduced $\sigma \equiv Z/(4 \pi a^{2})$, which is
henceforth referred to as the `\emph{surface charge density}' of
the droplet. The total charge of a planar `droplet' is infinite,
because of its infinite surface area. The results in
Table~\ref{tab:diff} show that the physical quantities $Z$,
$\gamma$, and $\Gamma$ are highly sensitive to the size $R$ of the
Wigner-Seitz cell in the experimentally relevant regime $R \approx
5 -10$ $\mu$m, $\epsilon_{o} \approx 5$, emphasizing the
importance of curvature. We will examine this dependence more
closely for extremely dilute emulsion, i.e., those emulsions which
can be modelled by a droplet in an infinitely large Wigner-Seitz
cell.

\section{Extremely dilute emulsions}

\subsection{\label{subsub:planar}Droplet size and curvature effects}

\begin{figure}
\includegraphics[width=3.375in]{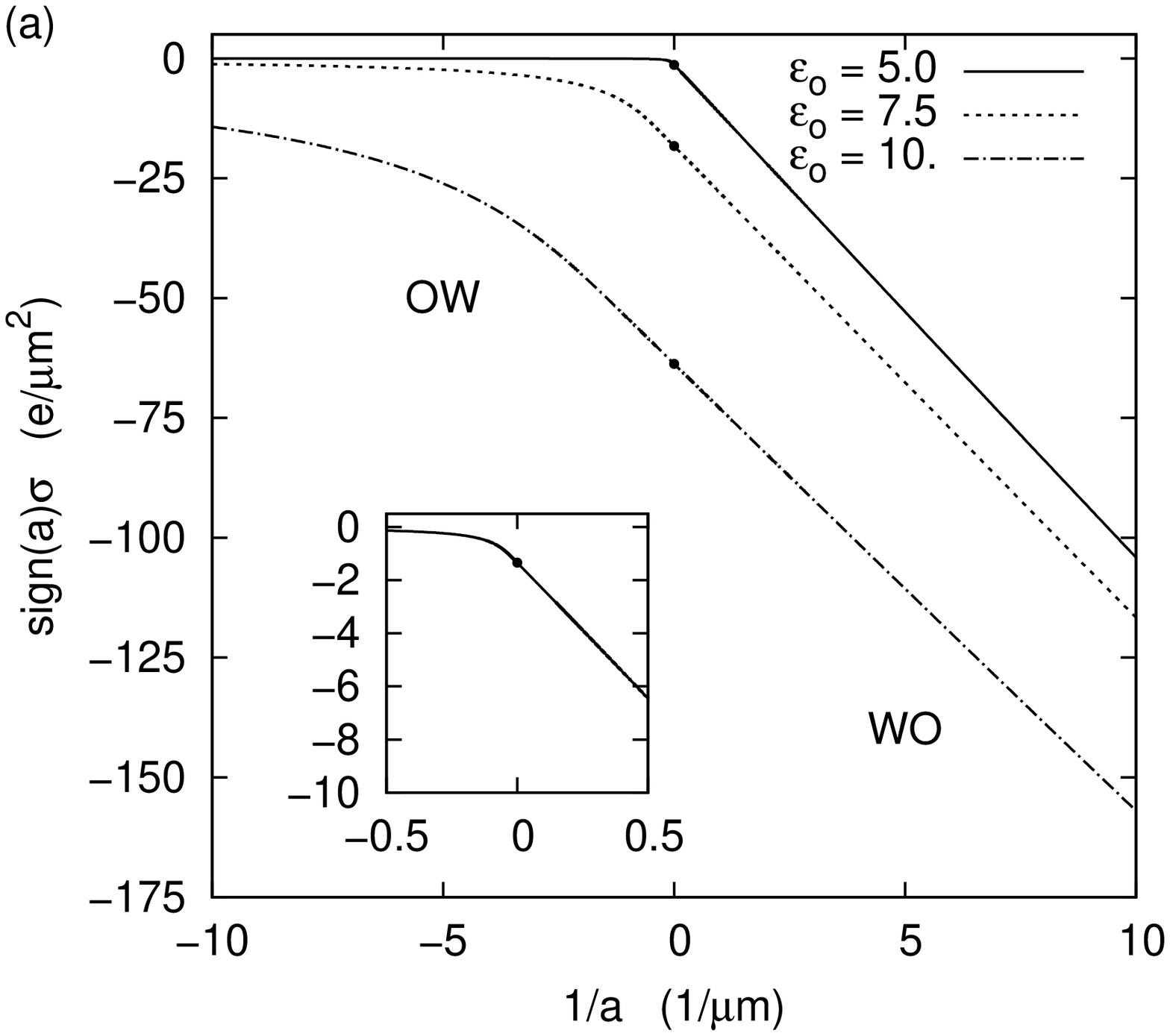}
\includegraphics[width=3.375in]{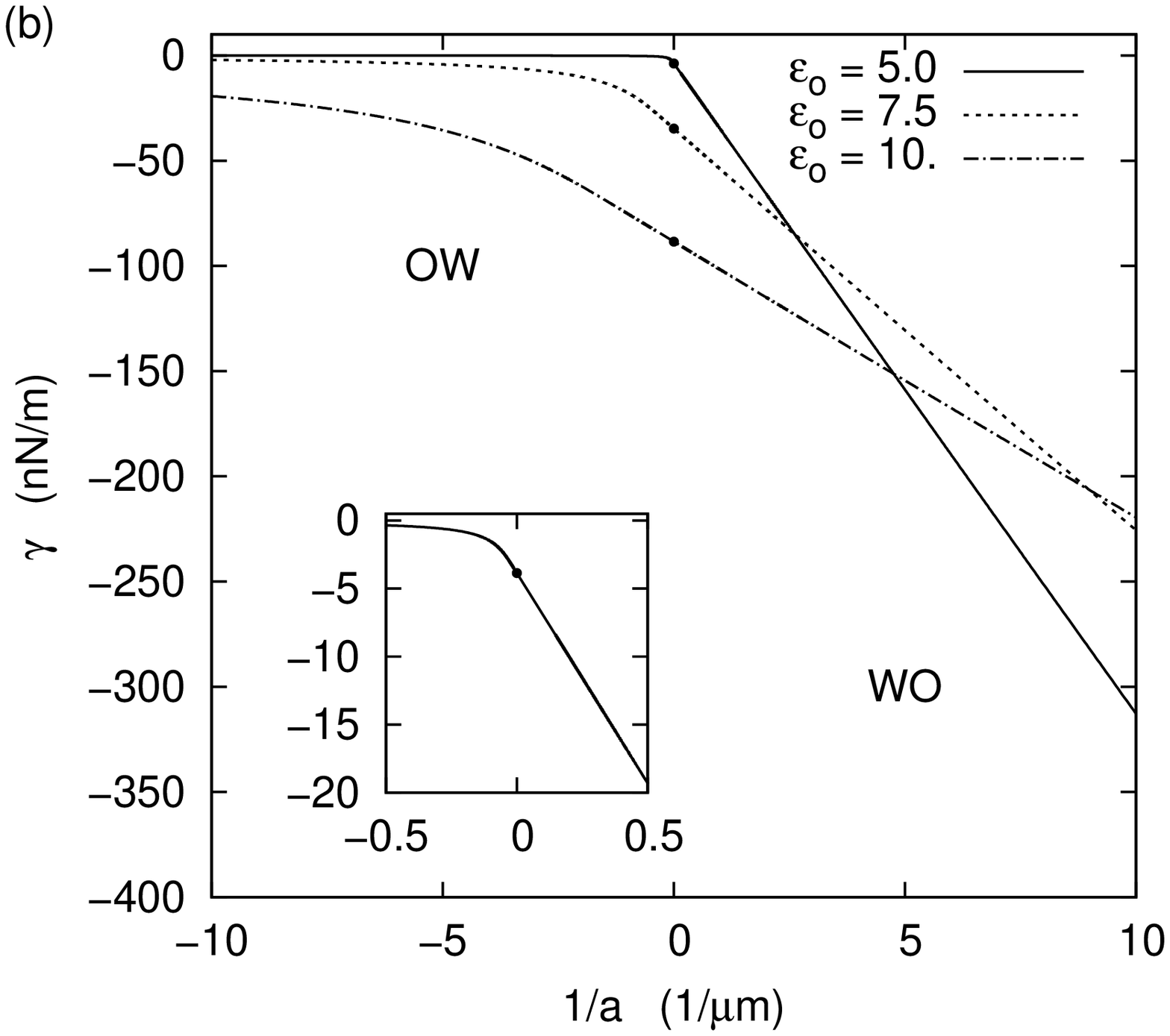}
\caption{\label{fig:planar} The surface charge density $\sigma$
(a) and the excess interfacial tension $\gamma$ (b) for an
extremely dilute system ($R \rightarrow \infty$) with salt
concentration in water $\rho_{w} = 10^{-3}$ M, as a function of
the water-droplet radius $a$ at several relative dielectric
constants $\epsilon_{o}$ of oil. As specified in
Fig.~\ref{fig:config}, positive values of $a$ correspond to WO
systems and negative values to OW systems. The inset shows a
enlargement of the area close to the origin. The analytic planar
limit values are indicated with dots. Note that the water-in-oil
droplets have negative charge, whereas the oil-in-water droplets
have positive charge.}
\end{figure}

We now disentangle the effects of a finite droplet density (finite
$R$) and that of droplet curvature (finite) $a$ by studying the
curvature dependence in the extremely dilute limit ($R \rightarrow
\infty$). In Fig.~\ref{fig:planar} the droplet's surface charge
density $\sigma$ and the excess interfacial tension $\gamma$ as a
function of $1/a$ are shown in the extremely dilute limit. For the
systems considered in Fig.~\ref{fig:planar}, it follows that there
is correspondence between the analytic planar results and the
limit $a \rightarrow \infty$, extrapolated from the spherical
results, within a fractional uncertainty of $\approx 10^{-4}$.
Data was obtained for sufficiently large $a$ to safely extend the
lines through the point $1/a = 0$. The surface charge density of
the systems considered here ($a > 100$ nm) lies in the $1 - 100$
$e/\mu\mathrm{m}^{2}$ range. We thus find that $a = 1$ $\mu$m
droplets have charges between $10 - 1000$ $e$. The excess
interfacial tension ranges between $10 - 100$ nN/m and is
negative. The sign for this interfacial tension is a consequence
of the external potential model we use here. We refer to
Ref.~\cite{mbier} for a more in-depth discussion of the excess
interfacial tension of an interface separating two electrolyte
solutions and the way in which it can be modelled.

The curves in Fig.~\ref{fig:planar} show that there are two
asymmetries between OW and WO emulsions. Firstly, the deviation
from the planar limit value is linear in $1/a$ in the case of WO
systems ($a > 0)$ and non-linear for OW systems ($a < 0$). This
asymmetry can be explained entirely by the fact that $\kappa_{w}a
\gg 1$ for the WO emulsions considered, whereas $\kappa_{o}a \ll
1$ for the OW emulsions. The deviation is linear for OW systems
only in the small regime near $1/a = 0$ where $\kappa_{o}a
> 1$. For $\kappa_{w}a \approx 1$ ($1/a \approx 100$ 1/$\mu$m) the deviation becomes non-linear
in WO emulsions as well, however, this is well beyond the scale
used in Fig.~\ref{fig:planar}. Secondly, with decreasing droplet
radius, $\vert \sigma \vert$ and $\vert \gamma \vert$ decrease for
OW systems, but these quantities increase for WO emulsions. This
effect can be attributed to the double-layer modification which
occurs in spherical systems. It can be shown that the oil phase
imposes the total structure of the double-layer and hence
determines the surface charge density and excess interfacial
tension. Curvature compresses the oil part of the double-layer in
OW emulsions, resulting in a reduction of $\vert \sigma \vert$ and
$\vert \gamma \vert$, whereas the oil part of the WO double-layer
gets stretched resulting in a corresponding increase of these
quantities. Note that the regime in which the planar limit
approximation gives accurate results for $\sigma$ and $\gamma$,
i.e., a deviation of less than say 20\%, is quite small. For an OW
emulsion with $\epsilon_{o} \lesssim 5$ the droplet's surface
charge density is negligible for experimentally reasonable droplet
radii~\cite{leuniss}. The planar value is an upper bound for the
quantities $\sigma$ and $\gamma$ in extremely dilute OW emulsion.
In WO emulsions, there may be an increase by 50\% (or sometimes
much more, see Table~\ref{tab:diff}) for $a = 1$ $\mu$m due to
curvature effects.

\subsection{\label{subsub:expan}Curvature expansions}

It is known for fluids with all intrinsic (correlation) length
scales smaller than all geometrical length scales that the
deviation of any intensive quantity from the planar value is a
linear combination of mean and Gaussian curvature
only~\cite{koenig}. In the spherical geometry considered here this
statement translates into a surface charge density as a function
of the droplet radius $a$ of the form
\begin{equation}
\label{eq:expans} \sigma(a) = \sigma_{p}\left( \textrm{sign}(a) +
\frac{c_{1}}{\vert \kappa_{o} a \vert} - \frac{c_{2}}{\vert
\kappa_{o} a \vert^{2}} \right),
\end{equation}
with $c_{1}$ and $c_{2}$ coefficients and $\sigma_{p}$ the
analytically known planar value for WO emulsions, \emph{provided}
$\kappa_{w}^{-1},\kappa_{o}^{-1} \ll a,R$. We use the sign
convention introduced in Fig.~\ref{fig:config}, which ensures that
Eq.~\cref{eq:expans} is valid for both OW and WO systems. A
similar expression can be found for $\gamma$, by replacing
$\sigma_p$ with $\mathrm{sign}(a)\gamma_p$. Moreover, $c_1$ and
$c_2$ are positive and for typical system parameters of order
unity. Equation~\cref{eq:expans} proves useful as it allows us to
describe the behavior of emulsion droplets for a range of droplet
radii by determining the charge and excess interfacial tension for
only two values of the droplet radius, fixing $c_{1}$ and $c_{2}$
in combination with the planar value. If the intrinsic and the
geometric length scales are \emph{not} well separated, higher
order terms in mean and Gaussian curvature can appear in
Eq.~\cref{eq:expans}~\cite{koenig}. However, these higher order
terms turn out to be small in this study, since
Eq.~\cref{eq:expans} quantitatively accounts for our numerical
data in the droplet radius regime $\vert a \vert > 0.1$ $\mu$m
even though $\kappa_{o}^{-1} \gtrsim a$ in part of this regime.
\begin{figure}
\includegraphics[width=3.375in]{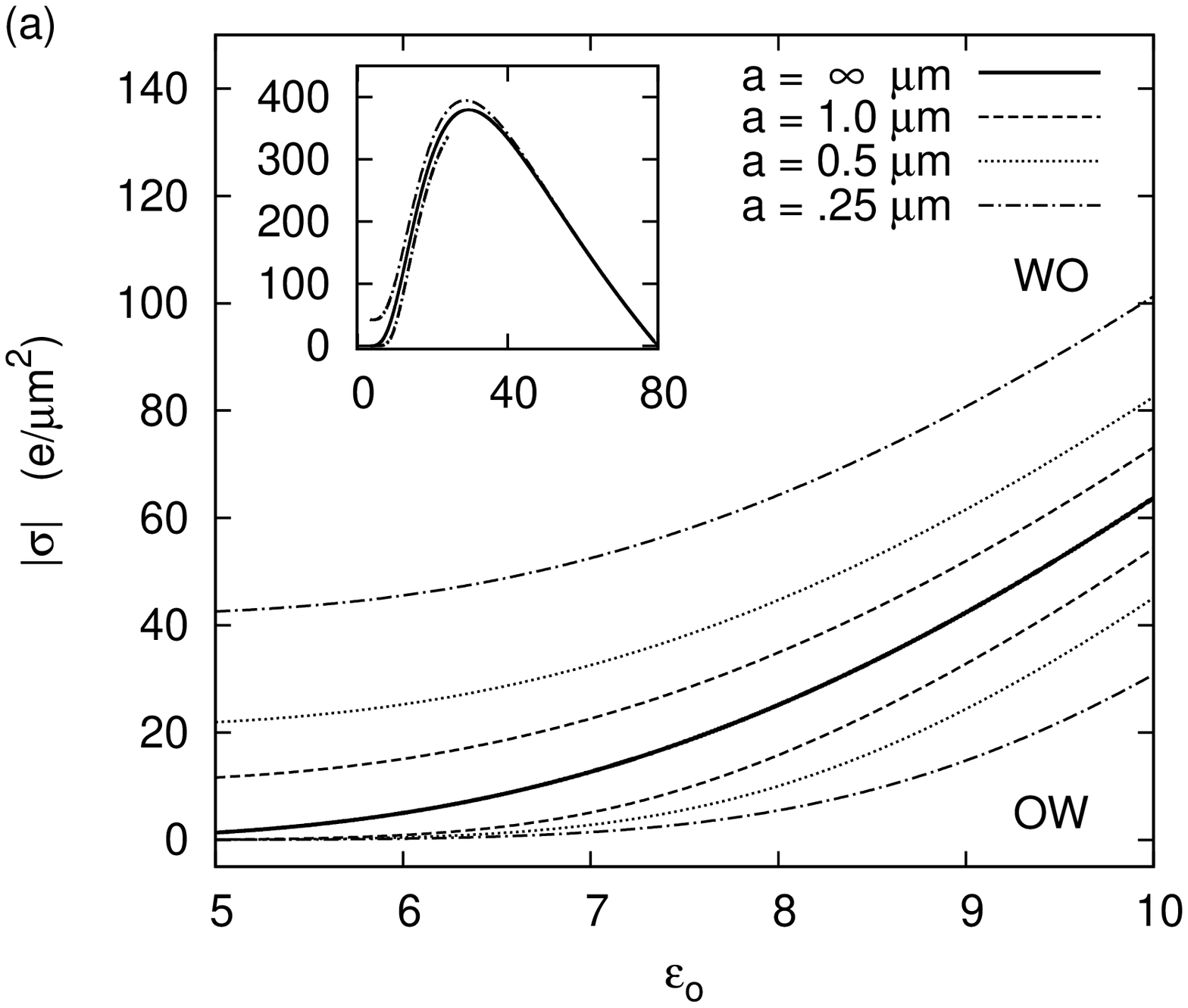}
\includegraphics[width=3.375in]{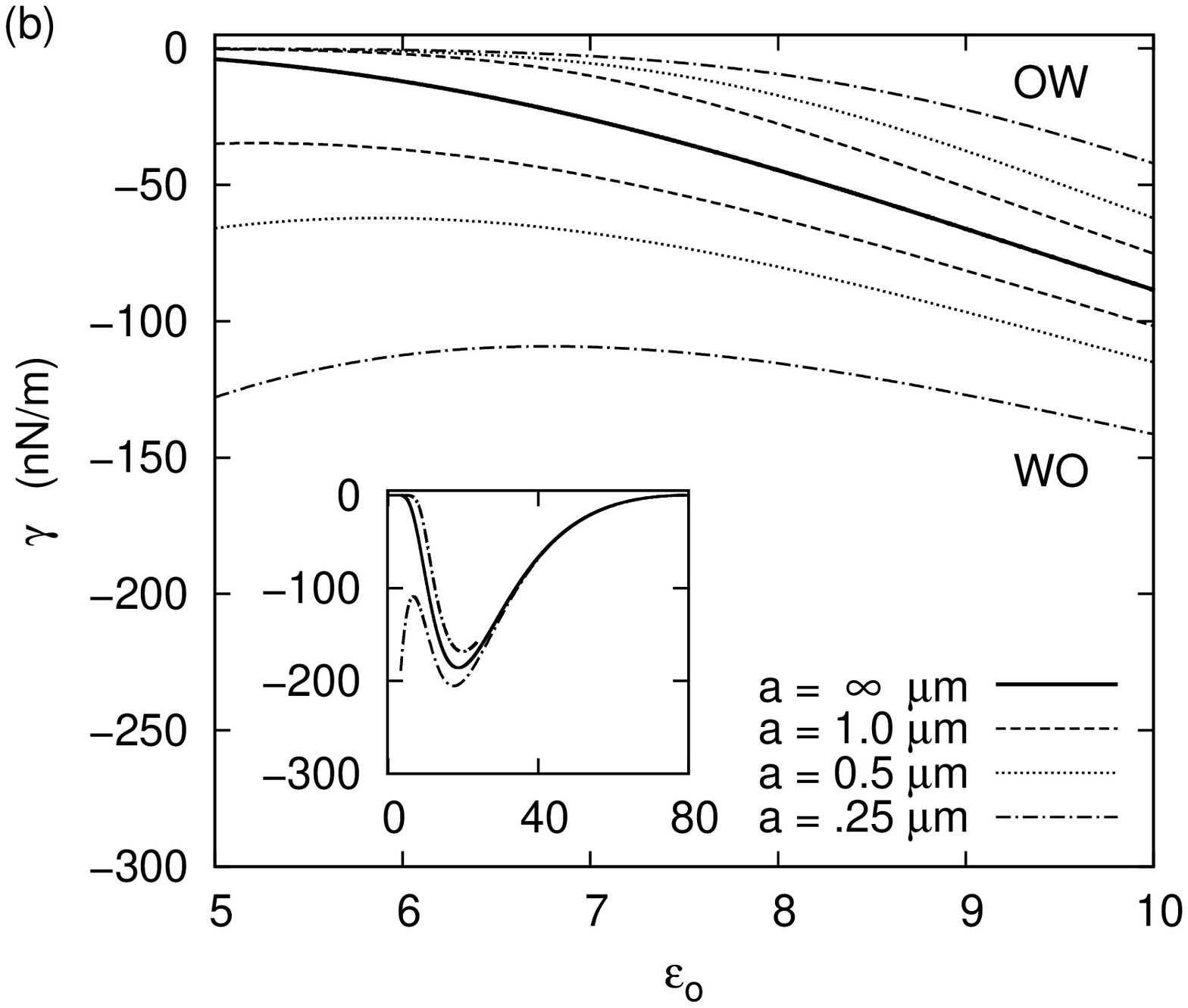}
\caption{\label{fig:epsilo} The surface charge density $\sigma$
(a) and the excess interfacial tension $\gamma$ (b) as a function
of the relative dielectric constant of oil $\epsilon_{o}$ for an
extremely dilute emulsion, with $\rho_{w} = 10^{-3}$ M. Several
droplet radii, $a = 0.25$, $0.5$, and $1.0$ $\mu$m, are compared
with the analytic planar values ($a \rightarrow \infty$). The
charge of the droplets in OW emulsions is positive and in WO
emulsions negative. The insets show the behavior of $\sigma$ and
$\gamma$ for the larger domain $3 < \epsilon_{o} < 80$, with both
the planar and the $a = 0.25$ $\mu$m result (OW and WO) indicated.
The OW line terminates due to numerical instabilities.}
\end{figure}

\subsection{\label{subsub:epsilo}Varying $\epsilon_{o}$, $\rho_{w}$ and the droplet radius}

The effect of double-layer modification on a droplet in an
extremely dilute system can also be evaluated when we vary
$\epsilon_{o}$ and $\rho_{w}$ in an experimentally reasonable
range. In Fig.~\ref{fig:epsilo}, $\sigma$ and $\gamma$ are given
as a function of $\epsilon_{o}$, with $R \rightarrow \infty$ and
$\rho_{w} = 10^{-3}$ M, for several droplet radii $a$. From the
insets, which show the full $\epsilon_{o}$-regime, we can see that
the planar limit approximation is very accurate in a large
$\epsilon_{o}$-range. However, it becomes apparent that in the
experimentally relevant $\epsilon_{o}$-range there is a
significant deviation from the planar value. In fact, for
$\epsilon_{o} < 7$ we see that OW systems hardly experience any
electrostatic effects ($\sigma \approx 0, \gamma \approx 0$),
whereas for WO emulsions such effects are much stronger than
planar theory predicts. We find that $\sigma$ is of the order $1 -
100$ $e/\mu\mathrm{m}^{2}$ and $\gamma$ is of order $10 - 100$
nN/m and negative.

Again there is an asymmetry between OW and WO systems, which can
be explained by the difference in Debye length w.r.t. the droplet
size. Note that the asymmetry between OW and WO becomes smaller
when the dielectric constant of the oil increases and hence the
Debye length in oil decreases. Between the two limiting cases
$\epsilon_{o} = 1$ and $\epsilon_{o} = \epsilon_{w}$, for which
both $\sigma$ and $\gamma$ are negligible in the former case and
must vanish in the latter case (in our model), we find an extremal
value for $\sigma$. This extremum is explained by an increase in
$\rho_{o}$, but a decrease in $f_{\pm}$ for $\epsilon_{o}
\rightarrow \epsilon_{w}$, and vice versa for $\epsilon_{o}
\rightarrow 1$. Note, however, that $\gamma$ seems to diverge for
the smallest water-in-oil droplet that we consider here ($a =
0.25$ $\mu$m) in the limit of small $\epsilon_{o}$ shown in the
inset of Fig.~\ref{fig:epsilo}b. A similar, non-vanishing behavior
can be observed for $\sigma$ in the inset of
Fig.~\ref{fig:epsilo}a, however, we do not find an apparent
divergence in $\sigma$ for the $\epsilon_{o}$-values considered
here. This is in contradiction with the intuitive idea that our
model should have negligible surface charge and excess surface
tension in the $\epsilon_{o} \rightarrow 1$ limit. Our theoretical
investigation cannot exclude instabilities and uncertainties in
the numerical algorithm used to solve for $\phi(r)$ in the extreme
$\epsilon_{o}$-regime. A more detailed evaluation of the limiting
behavior for WO and OW emulsions with $\epsilon_{0} \approx 1$
(e.g., water-droplets in air and air-bubbles in water
respectively) with a spherical interface is beyond the scope of
this work.

\begin{figure}
\includegraphics[width=3.375in]{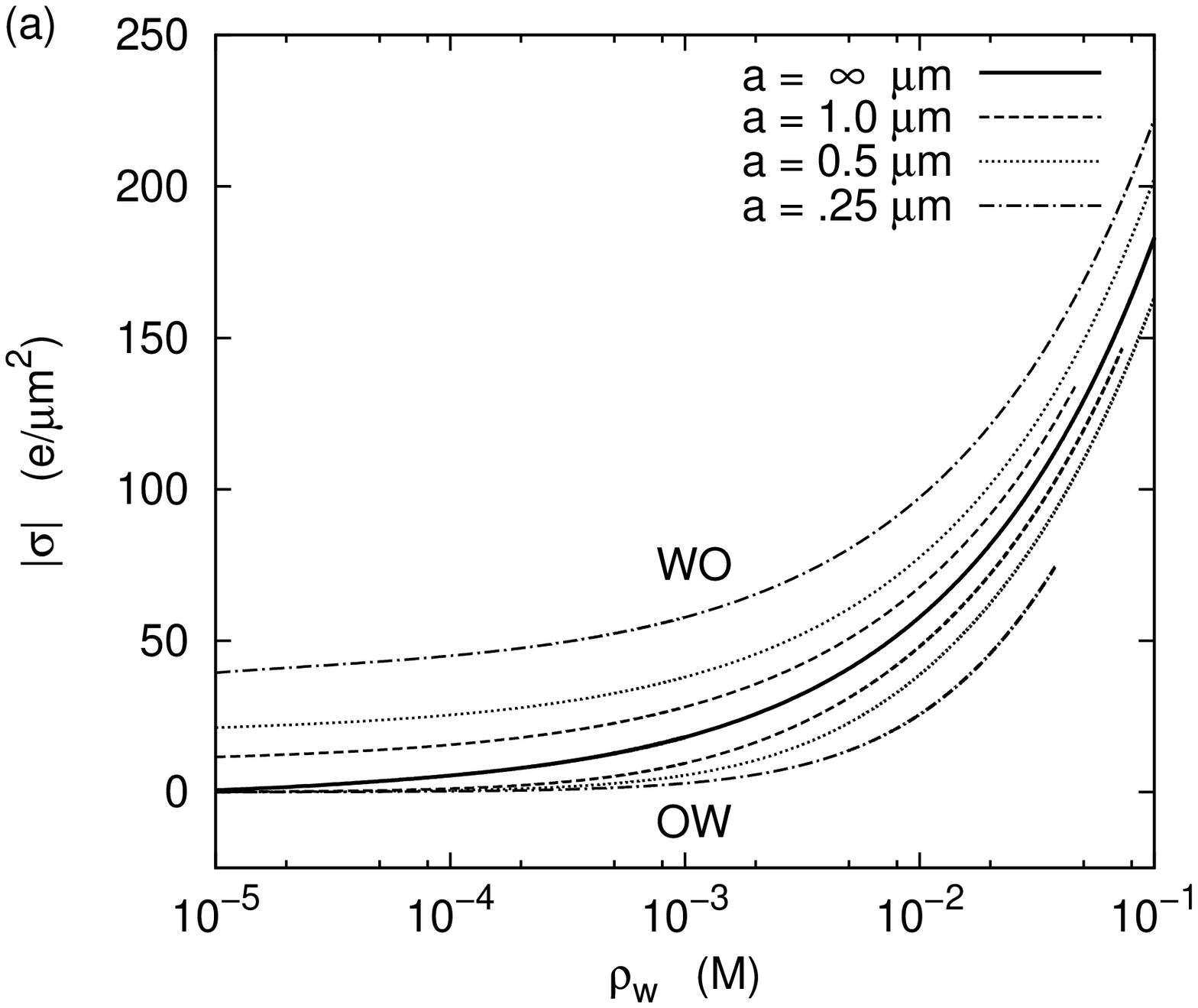}
\includegraphics[width=3.375in]{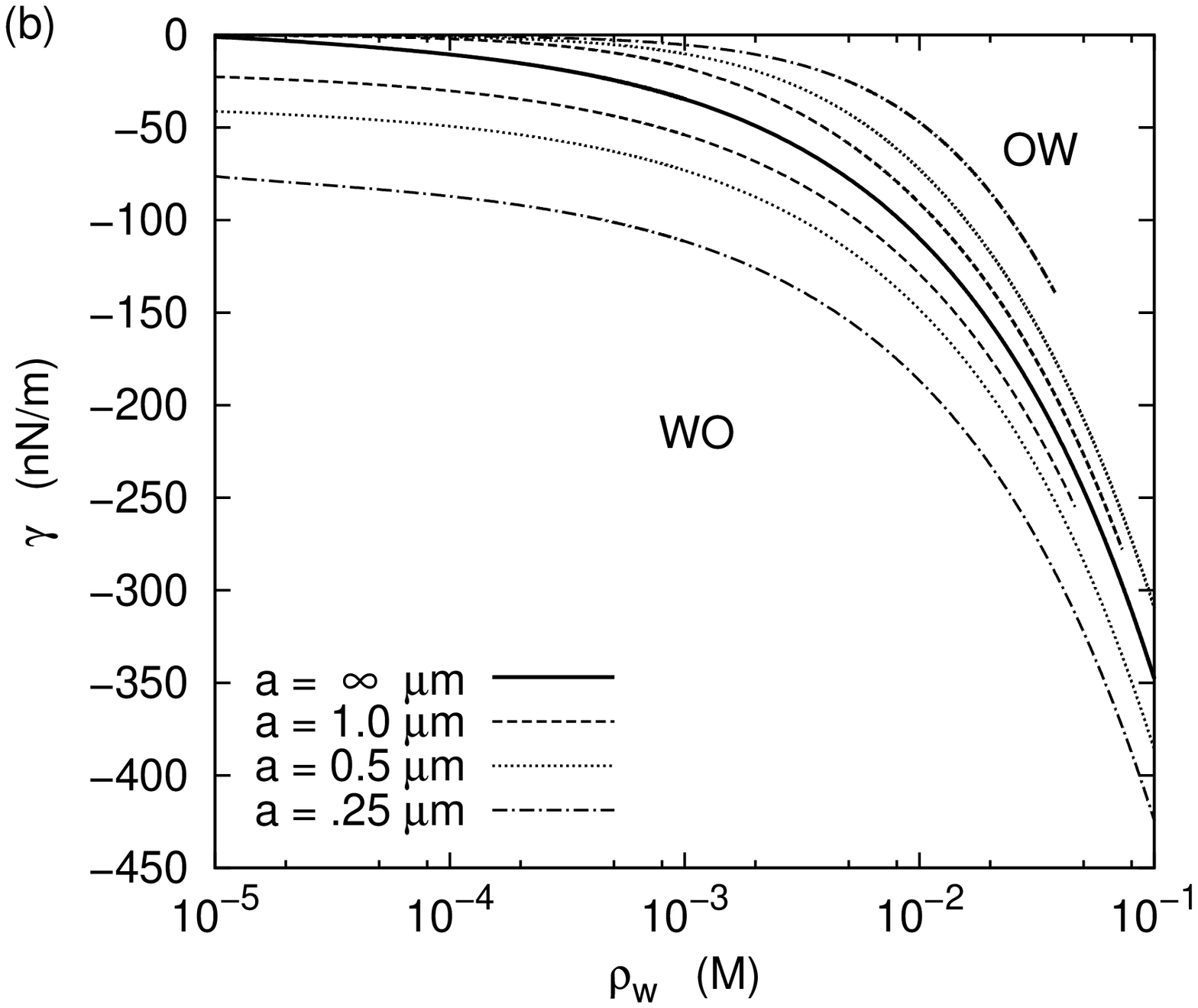}
\caption{\label{fig:densit} The surface charge density $\sigma$
(a) and the excess interfacial tension $\gamma$ (b) as a function
of the bulk ion concentration in water $\rho_{w}$ for an extremely
dilute emulsion with $\epsilon_{o} = 7.5$. Several droplet radii,
$a = 0.25$, $0.5$, and $1.0$ $\mu$m, are considered and can be
compared to the planar value. The charge of the droplets in OW
emulsions is positive and in WO emulsions negative. Some lines
terminate due to numerical instabilities.}
\end{figure}

In Fig.~\ref{fig:densit} we show $\sigma$ and $\gamma$ as a
function of $\rho_{w}$ in an extremely dilute system with
$\epsilon_{o} = 7.5$ for several droplet radii $a$. One can see
that the behavior of the physical quantities for spherical
interfaces w.r.t. their planar counterparts is analogous to that
found in Fig.~\ref{fig:epsilo}. This analogy can be easily
explained by the way in which the Debye lengths are modified when
changing either $\epsilon_{o}$ or $\rho_{w}$. Note the values of
$\sigma$ and $\gamma$ are in the range $1 - 100$
$e/\mu\mathrm{m}^{2}$ and $10 - 100$ nN/m, respectively. Our
results for extremely dilute emulsions,
Figs.~\ref{fig:planar},~\ref{fig:epsilo}, and~\ref{fig:densit},
thus show that for reasonable choices of the system's parameters
$\vert \gamma \vert$ is of the order $10 - 100$ nN/m. This excess
interfacial tension therefore does not contribute significantly to
the bare interfacial tension of an oil-water interface, which is
of the order $1 - 10$ mN/m.

\section{Crystallization at finite droplet volume fraction}

\begin{figure}
\includegraphics[width=3.375in]{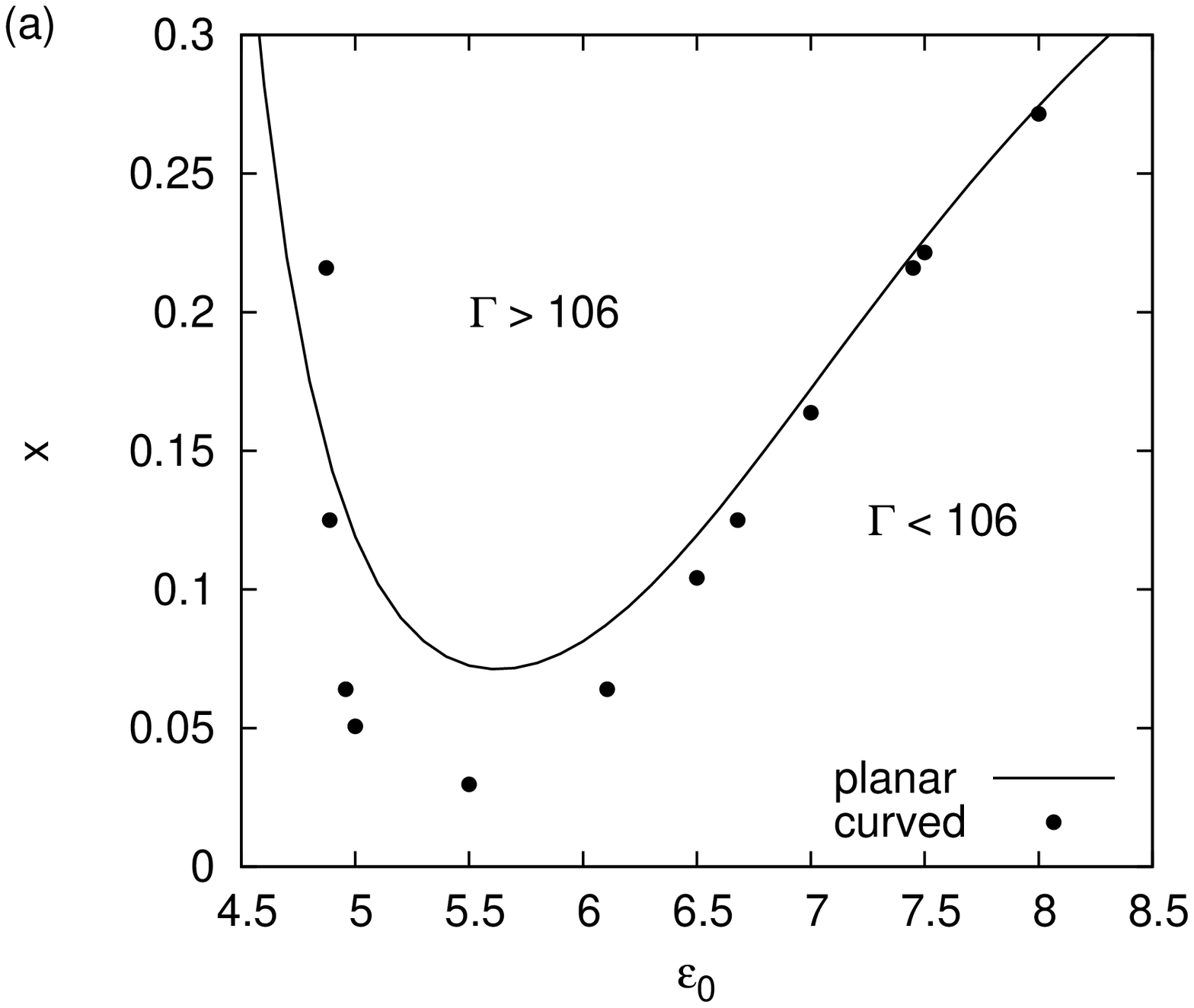}
\includegraphics[width=3.375in]{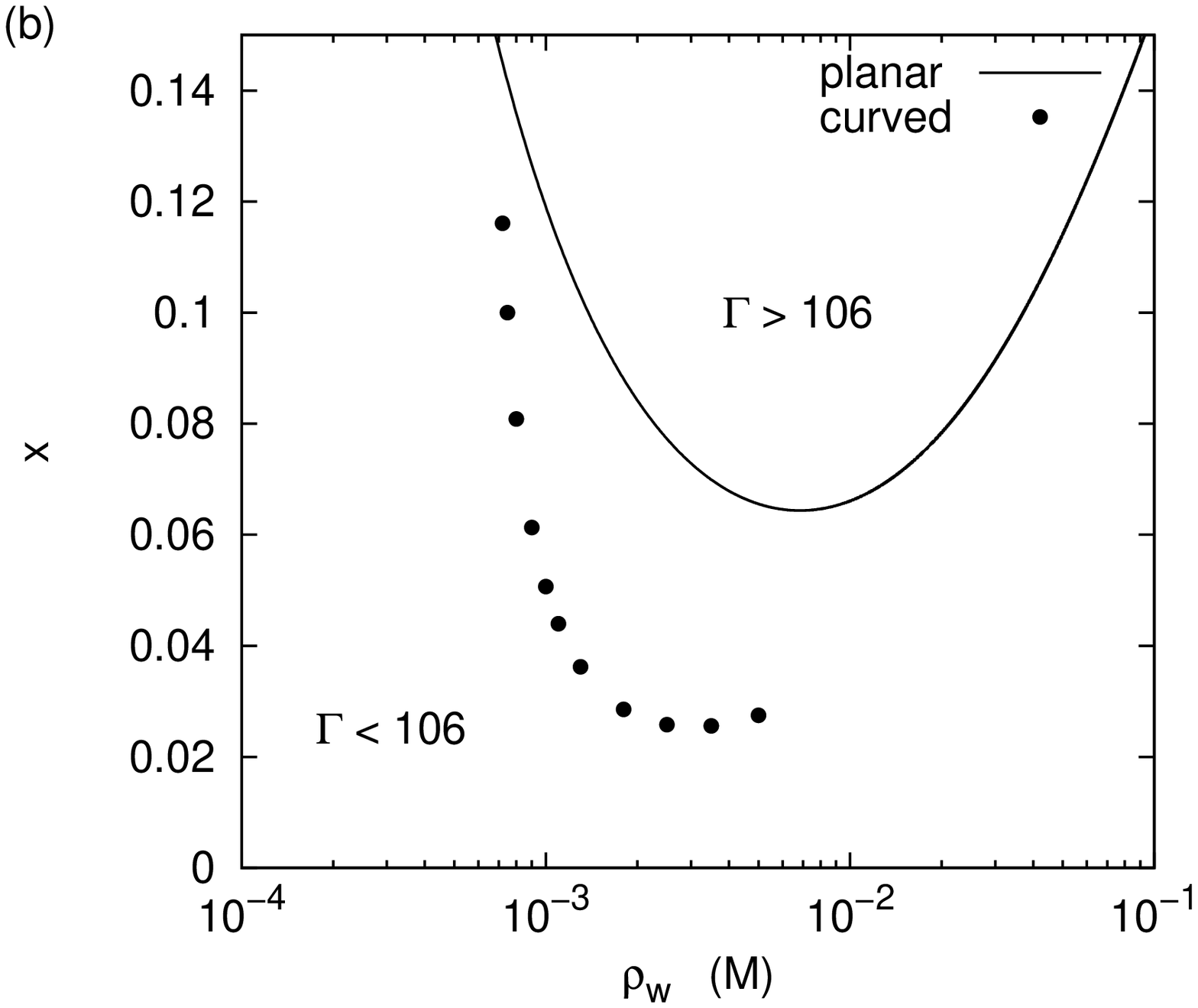}
\caption{\label{fig:isolin} The freezing line, $\Gamma = 106$, as
a function of the relative dielectric constant of oil
$\epsilon_{o}$ and the water volume fraction $x$ for $R = 10$
$\mu$m and $\rho_{w} = 10^{-3}$ M (a) and as a function of
$\rho_{w}$ and $x$ for $R = 10$ $\mu$m and $\epsilon_{o} = 5$ (b).
The planar $\Gamma = 106$ isoline is also indicated and several
data points which lie on the spherical interface freezing line are
included. Above the isolines $\Gamma > 106$ and below $\Gamma <
106$. Note the shift in the freezing line minima and the increase
in the crystallization zones.}
\end{figure}

In this section we consider a Wigner-Seitz cell with $R = 10$
$\mu$m, yielding a bcc nearest neighbor distance $l =
(\sqrt{3}\pi)^{1/3}R = 17.6$ $\mu$m, which for $\epsilon_{o}
\approx 5$ is within the regime for the experiments of
Refs.~\cite{leuniss,miriaml}. Only WO emulsions are examined, as
we found no crystallization for OW emulsions, i.e., $\Gamma \ll
106$ (see Eq.~\cref{eq:GAMMA}) for any reasonable choices of the
OW system parameters. Our model thus predicts that crystallization
of oil-in-water droplets does not occur or is extremely unlikely.
Figure~\ref{fig:isolin} indicates for WO systems the $\Gamma =
106$ isoline, which is in fact the WO droplet \emph{freezing
line}, in the ($\epsilon_{o}$,$x$)-plane (a) and the
($\rho_{w}$,$x$)-plane (b) for our basis parameter set. The choice
for the $\epsilon_{o}$-range in Fig.~\ref{fig:isolin}a is inspired
by the range of dielectric constants for which droplet crystal
formation has been observed~\cite{leuniss}. The $\rho_{w}$-range
in Fig.~\ref{fig:isolin}b is physically reasonable, inspired by
the isoline minimum found in planar analysis, and limited by the
stability of our numerical algorithm to solve for $\phi(r)$ in the
spherical geometry.

We see that there is an isoline minimum at $\epsilon_{0} \approx
5.5$ ($\rho_{w} = 10^{-3}$ M) and at $\rho_{w} \approx
3.0\cdot10^{-3}$ M ($\epsilon_{0} = 5$) with $x \approx 0.025$ for
the spherical results. This value is substantially smaller than
that of the planar minimum ($x \approx 0.065$), however it is
still significantly larger than the experimentally found water
content of emulsions in which water-in-oil crystals were observed.
This was to be expected if one considers the uncertainty of some
of the parameters used, particularly the ionic contents of the
emulsions and the corresponding ionic self-energies. The location
of the minimum also gives an a posteriori justification of our
choice to use $\rho_{w} = 10^{-3}$ M for our basis parameter set.
Note the regime in which crystallization can occur according to
spherical theory is greatly extended with respect to that found
using planar theory. Yet there are parameter choices for which
this regime is reduced, also see Fig.~\ref{fig:isolin}. Therefore,
we must conclude that the effects of curvature on the
crystallization of spherical water-in-oil droplets are non-trivial
and involve competing processes: at extreme dilution the surface
charge density is higher than the planar value, but at finite
concentration it becomes smaller, also see Table~\ref{tab:diff}.
In accordance with the rough location of the minimum found in
Fig.~\ref{fig:isolin}, we will use $\epsilon_{o} = 5$ and
$\rho_{w} = 10^{-3}$ M to examine the effects of the self-energy
difference between water and oil of the respective ions, keeping
$R = 10$ $\mu$m ($l = 17.6$ $\mu$m) fixed.

\begin{figure}
\includegraphics[width=3.375in]{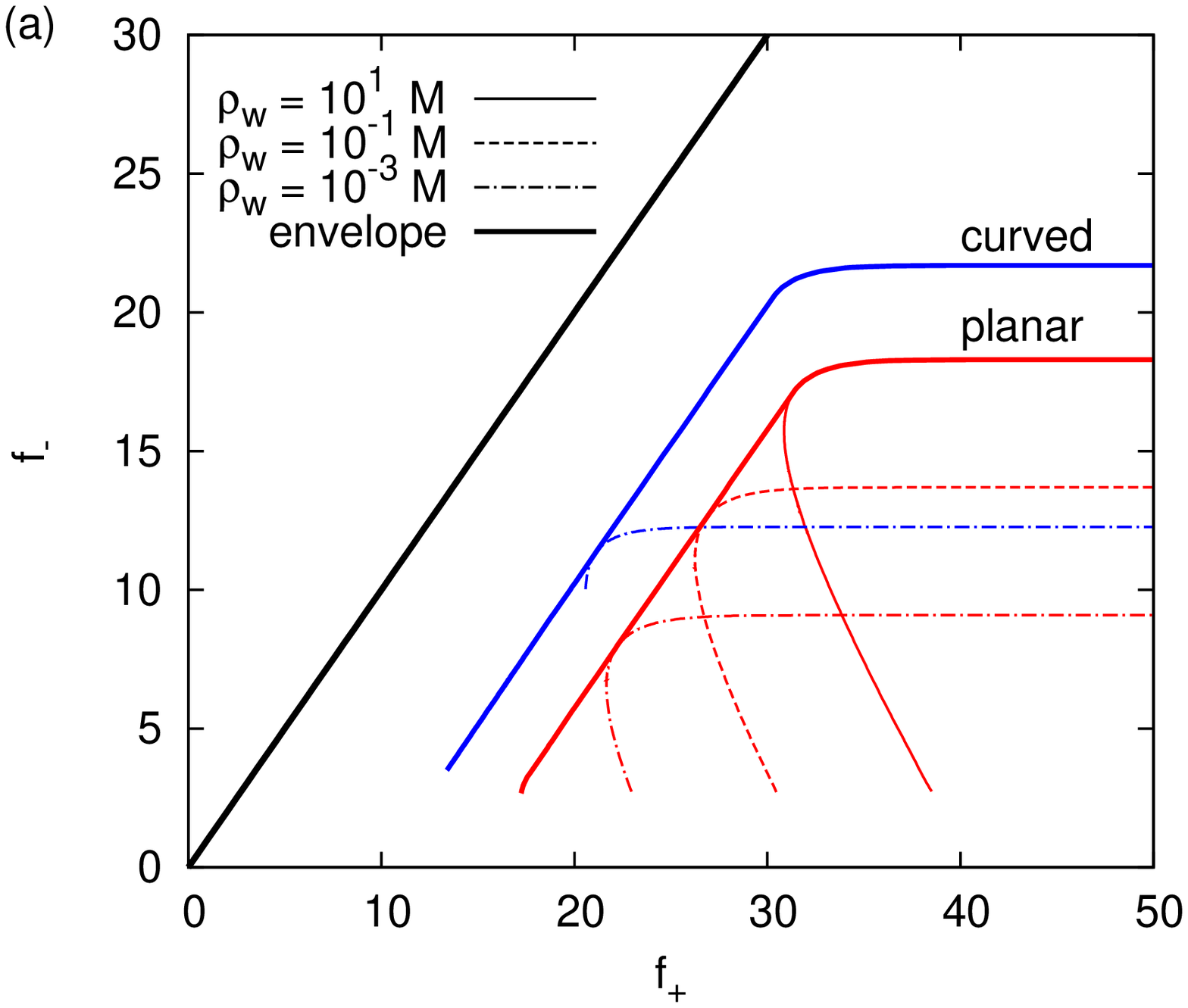}
\includegraphics[width=3.375in]{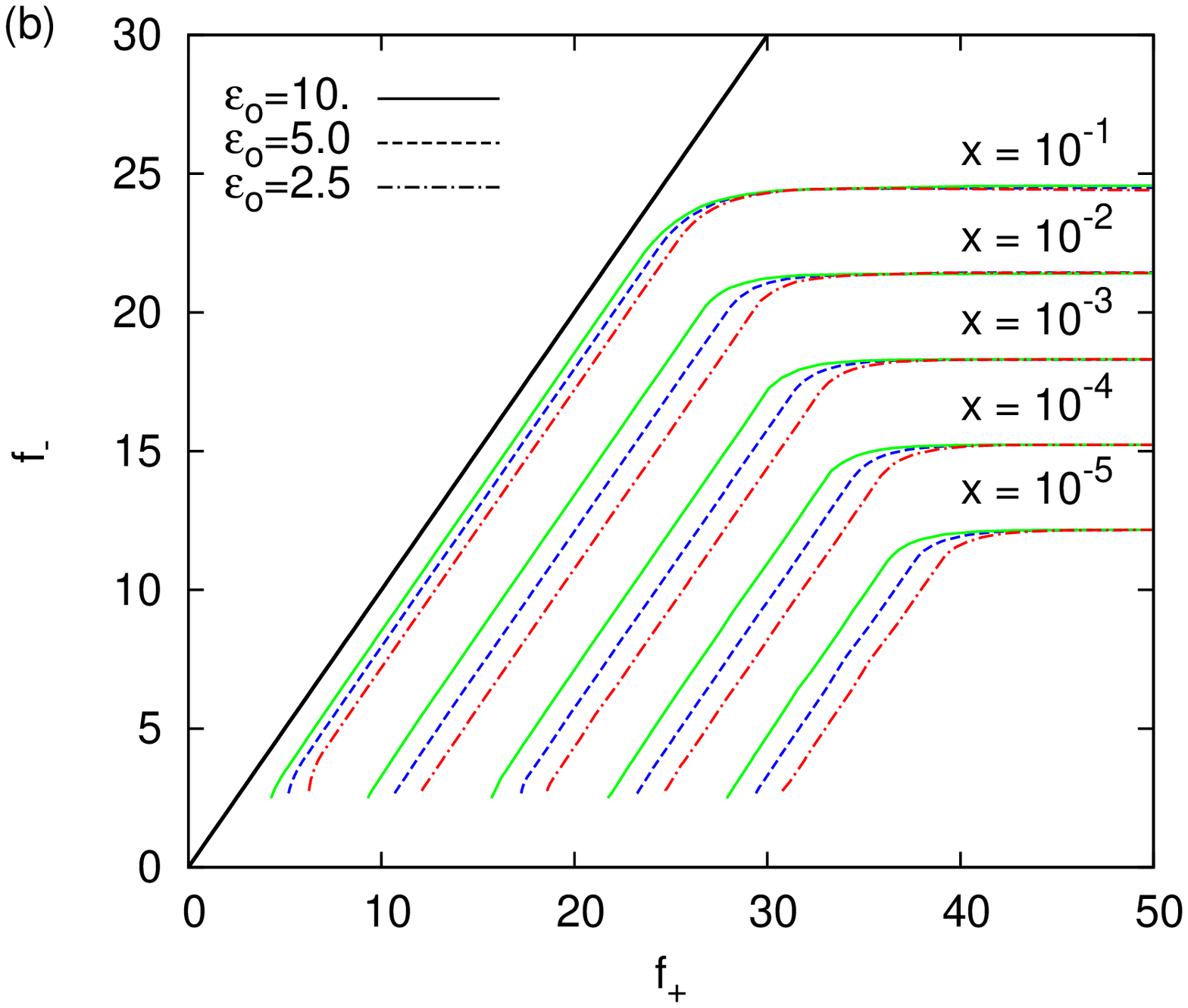}
\caption{\label{fig:envel} (Color online) Coupling parameter
isolines $\Gamma = 106$ for water-in-oil droplets in the
$(f_{+},f_{-})$ plane (see text), in (a) for several salt
concentrations $\rho_{w}$ (dashed) and the $\rho_{w}$-envelopes
(full curves) based on both curved and planar geometries at
composition $x = 10^{-3}$ and oil dielectric constant
$\epsilon_{o} = 5$, in (b) only the $\rho_w$-envelopes at several
$x$ and $\epsilon_{o}$ based on the planar geometry. In all cases
the droplet density is kept fixed such that $R = 10$ $\mu$m. The
diagonal ($f_{+} = f_{-}$) is a reflection symmetry axis for all
curves. Crystallization is predicted between the $f_{+}$ axis and
the curves (and between their reflected images of course).}
\end{figure}

In Fig.~\ref{fig:envel}a the $\Gamma = 106$ isolines are indicated
as a function of $f_{\pm}$ for several $\rho_{w}$ and $x =
10^{-3}$, together with the convex envelope of the isolines within
the $\rho_{w} = 10^{-7} - 10$ M range for both spherical and
planar interfaces. The choice of this $\rho_{w}$ envelope range is
inspired by well-known numbers for the ion concentration in water,
which is bounded from below by that of pure water with a pH of 7
caused by self-dissociation of water molecules and from above by
that of a saturated solution of order 10 M. Note that
crystallization of water-droplets in oil can occur in the region
enclosed by the isolines or envelopes, respectively, and the
$f_{+}$-axis, i.e., in this region $\Gamma > 106$.
The spherical interface envelope in Fig.~\ref{fig:envel}a was
determined by calculating the $\Gamma = 106$ isolines for
$\rho_{w} = 10^{-6} - 10^{-2}$ M. All of these isolines turn out to be
shifted in the same way with respect to their planar counterparts
as is shown explicitly for $\rho_{w} = 10^{-3}$ M. Hence, we have assumed
that this behavior can be extrapolated to $\rho_{w} = 10$ M. Note
again that the effect of curvature is to increase the range over
which crystallization can occur.

Fig.~\ref{fig:envel}b shows these convex envelopes, in the planar
limit approximation, as a function of $f_{\pm}$ for several $x$
and $\epsilon_{o}$. In agreement with our findings in
Fig.~\ref{fig:isolin}, we recover that crystallization occurs more
easily for larger $x$, as one would expect, since larger droplets
have a higher charge and are closer together when $R$ is fixed.
Note that in the 2.5 - 10 $\epsilon_{o}$-range there is no
significant shift in the convex envelopes and their corresponding
$\Gamma = 106$ isolines. The spherical interface results have not
been considered here, because of the time consuming character of
these calculations. However, one can expect an increase in the
crystallization zone for these envelopes similar to that of
Fig.~\ref{fig:envel}a.

\section{\label{sec:concl}Conclusion and Outlook}

We have presented calculations for anions and cations near a
spherical water-oil interface, taking into account ionic
self-energies and screening, to describe the spontaneous charging
of water droplets in oil. This theory was applied to emulsions of
oil and water which contain ions.

In the extremely dilute droplet limit, the effects of curvature on
the charge and excess interfacial tension induced by ion
partitioning of the anions and cations were compared to results
obtained for a planar interface. It turns out that the planar
limit approximation used in Ref.~\cite{zwanikk} can be applied
with a high degree of accuracy for many system parameters.
However, in the range of the experiments of
Refs.~\cite{leuniss,miriaml}, we have shown that spherical and
planar results differ significantly. Water-in-oil droplets have a
substantially higher and oil-in-water droplets a substantially
lower charge/excess interfacial tension than one would expect on
the basis of planar calculations. In accordance with
Ref.~\cite{koenig} we found that the value of physical quantities
in a spherical system can to an extent be approximated using a
polynomial expansion in $1/\vert \kappa_{o}a \vert$ around the
planar value.

For finite volume fractions of water in oil we have investigated
the crystallization of water-in-oil droplets using the
dimensionless coupling parameter $\Gamma$ of a point-Yukawa system
to predict crystallization~\cite{plsmpr1}. The range in parameter
space in which crystal formation can occur is greatly extended by
using spherical values w.r.t. the planar result, mainly because of
the larger surface charge densities in the spherical case. We
expect that the theory we have presented captures the physics of
the experiments performed by Leunissen \emph{et al}. in
Refs.~\cite{leuniss,miriaml}. However, quantitative comparison
between theory and experiment is not possible at this time. Not
only additional theoretical effort is required, e.g., including
more realistic self-energies~\cite{netz} or a wider class
thereof~\cite{onuki0,onuki1,onuki2}, but the complex chemistry in
these oil-water emulsions needs to be further scrutinized and
elucidated experimentally to facilitate such a comparison.

Another extension of the present theory for ions in the vicinity
of a curved oil-water interface is the addition of charged
colloidal particles. It was shown
experimentally~\cite{leuniss,miriaml} and theoretically in the
planar geometry~\cite{miriaml,zwanikk} that the phenomenology due
to the presence of charged colloids is extremely rich, e.g.
involving strong colloidal adsorption at the droplet surface with
adjacent huge colloid-free zones. One would expect curvature
effects in these systems as well. Moreover, the much smaller
oil-in-water droplets in the $10-100$ nm range as observed in the
Pickering emulsions of Refs.~\cite{stapick,sponemu} certainly
warrant a theoretical treatment that takes the finite curvature of
the droplets into account. Studies along these lines are in
progress.

\section{\label{sec:ackn}Acknowledgements}

It is a pleasure to thank M. E. Leunissen and A. van Blaaderen for
useful discussions and sharing unpublished experimental data with
us. This work is part of the research program of the `Stichting
voor Fundamenteel Onderzoek der Materie (FOM)', which is
financially supported by the `Nederlandse Organisatie voor
Wetenschappelijk Onderzoek (NWO)'.

\bibliography{BIBL}

\end{document}